\documentclass[aps,twocolumn,showpacs,preprintnumbers,amsmath,amssymb]{revtex4}

\usepackage{graphicx}
\usepackage{dcolumn}
\usepackage{bm}

\begin{document}
\newlength{\figwidth}
\setlength{\figwidth}{0.5 \textwidth}
\addtolength{\figwidth}{-0.5 \columnsep}
\addtolength{\figwidth}{-1cm}

\preprint{}

\title{Random phase vector 
for calculating the trace of a large matrix}

\author{Toshiaki Iitaka}
 \email{tiitaka@riken.jp}  
 \homepage{http://www.iitaka.org/}
\author{Toshikazu Ebisuzaki}%
\affiliation{
Computational Astrophysics Laboratory, \\
RIKEN (The Institute of Physical and Chemical Research) \\
2-1 Hirosawa, Wako, Saitama 351-0198, Japan} 

\date{\today}

\begin{abstract}
We derive an estimate of statistical error in calculating the trace of a large matrix by
using random vector, and show that {\em random phase vector} gives the results with the smallest
statistical error for a given basis set.
This result supports use of random phase vectors in the calculation of density of states and
linear response functions of large quantum systems.
\end{abstract}

\maketitle

In fast algorithm called {\em Order-N methods} for calculating DOS and linear response functions \cite{Skilling1989,Drabold1993,Silver1994,Wang1994a,Wang1994b,Sankey1994,Alben1975,Iitaka1995b,Iitaka1997,Hams2000,Gelman2003,Nomura1997,Nomura1999,Iitaka1999,Iitaka2003,DeVries1993}, Monte Carlo calculation of the trace of large matrices by using {\em random vector} often plays an important role. The central limit theorem guarantees the convergence of the result as the sample number $K$ increases. Further experience shows another useful feature called {\em self-averaging effect}: the fluctuation in some physical quantities such as energy density and linear response functions decreases as the dimension $N$ of the Hilbert space increases\cite{Iitaka1997,Hams2000,Gelman2003,Nomura1997,Nomura1999,Iitaka1999,Iitaka2003,DeVries1993}. These two types of convergence make the random vector a very efficient numerical tool.

A special class of random vector called {\em random phase vector} \cite{Alben1975,Iitaka1995b,Iitaka1997}  has been used in later papers \cite{Nomura1997,Nomura1999,Iitaka1999,Iitaka2003,Gelman2003} without examining its efficiency rigorously. 
In this article, we prove , by following the scheme of \cite{Hams2000}, that random phase vector is really the most efficient random vector in a wide class of random vectors. Then we illustrate the mechanism of self-averaging with a simple model Hamiltonian.

A {\it (complex) random vector} is defined by 
\begin{equation}
\label{random_eq:randomvec}
|\Phi \rangle \equiv \sum_{n=1}^N |n \rangle \xi_n 
\end{equation}
where $\{ |n \rangle \}$ is the basis set used in the computation and 
$\xi_n$ are a set of complex random variables with identical probability distribution satisfying the statistical relations 
\begin{eqnarray}
\label{eq:statistical1}
\left\langle \left\langle \   \xi_{n} \  \right\rangle \right\rangle &=& 0 \\
\label{eq:statistical2}
\left\langle \left\langle \  \xi_{n_1}  \xi_{n_2} \  \right\rangle \right\rangle &=& 0 \\
\label{eq:statistical3}
\left\langle \left\langle \  \xi_{n_1}^*  \xi_{n_2} \  \right\rangle \right\rangle &=& \delta_{n_1n_2}
\end{eqnarray}
where $\left\langle \left\langle \   \cdot \  \right\rangle \right\rangle \ $ stands for the statistical average. 
This class of random vectors defined in one orthonormal basis set have the coefficients $\zeta_l$ in another orthonormal basis set $|l\rangle$ (for example, energy eigenstates $|E_l\rangle$ of a Hamiltonian $H$),
\begin{equation}
|\Phi \rangle = \sum_{l=1}^N |l \rangle \zeta_l
\end{equation}
that satisfy the same statistical relations as (\ref{eq:statistical1})-(\ref{eq:statistical3}):
Since the coefficients in the two basis sets are related by unitary transformation
\begin{eqnarray}
\label{random_eq:unitary}
\zeta_l     &=& \sum_{n=1}^N \langle l | n \rangle \xi_n \\
\zeta_l^{*} &=& \sum_{n=1}^N  \xi_n^{*} \langle n | l \rangle 
,
\end{eqnarray}
the statistical relations of $\zeta_n$ are derived as
\begin{eqnarray}
\left\langle \left\langle \   \zeta_{l} \  \right\rangle \right\rangle 
&=& \sum_{n=1}^N 
  \langle l |n \rangle
\left\langle \left\langle \   \xi_{n} \  \right\rangle \right\rangle 
= 0 \\
\left\langle \left\langle \  \zeta_{l_1}  \zeta_{l_2} \  \right\rangle \right\rangle 
&=& \sum_{l_1=1}^N  \sum_{l_2=1}^N 
\langle l_1 |n_1 \rangle \langle l_2 | n_2 \rangle
\left\langle \left\langle \  \xi_{n_1}  \xi_{n_2} \  \right\rangle \right\rangle = 0 \\
\label{random_eq:stat.zeta}
\left\langle \left\langle \  \zeta_{l_1}^*  \zeta_{l_2} \  \right\rangle \right\rangle 
&=& \sum_{l_1=1}^N  \sum_{l_2=1}^N 
\langle n_1 |l_1 \rangle \langle l_2 | n_2 \rangle
\left\langle \left\langle \  \xi_{n_1}^*  \xi_{n_2} \  \right\rangle \right\rangle \nonumber \\
&=& \sum_{n=1}^N \langle l_2 | n \rangle \langle n | l_1 \rangle 
= \langle l_2 | l_1 \rangle 
= \delta_{l_1l_2}.
\end{eqnarray}
In view of energy eigenstates, uniform random vector contains all eigenstates with equal probability and represents the system at a very high temperature.  
The {\it orthonormality} and {\it completeness} of random vectors
\begin{eqnarray}
\label{random_eq:randomvector.normal}
\langle \langle \ \ \ \langle \Phi | \Phi \rangle \ \ \ \rangle \rangle &=& N \\
\label{random_eq:randomvector.ortho}
\langle \langle \ \ \ | \Phi \rangle \langle \Phi | \ \ \  \rangle \rangle &=& {\bf I}
\end{eqnarray}
where $\bf I$ is the identity operator, are shown by using (\ref{eq:statistical1})-(\ref{eq:statistical3}).

The most important feature of random vectors is that statistical average of $\langle \Phi | X | \Phi \rangle$ gives the trace of $X$ as follows:
\begin{eqnarray}
\label{random_eq:trace.monte.basis}
&& \left\langle \left\langle \   \langle \Phi | X | \Phi \rangle  \  \right\rangle \right\rangle \nonumber \\  
&=& \sum_{n}
 X_{nn}  
+ \sum_{ n_1, n_2} \left\langle \left\langle \ \xi_{n_1}^* \xi_{n_2}  -\delta_{n_1n_2} \  \right\rangle \right\rangle \  X_{n_1,n_2} \\
&=& \sum_{ n } X_{nn} = {\rm tr} \left[ X \right] \nonumber
.
\end{eqnarray}
For numerical evaluation of (\ref{random_eq:trace.monte.basis}) with $K$ samples of random vectors, the second term of (\ref{random_eq:trace.monte.basis}) gives the statistical fluctuation,
\begin{equation}
\label{eq:fluctuation0}
\delta X = \frac{1}{K} \sum_{k}\sum_{ n_1, n_2} \left( \xi_{n_1}^{*(k)} \xi_{n_2}^{(k)}  -\delta_{n_1n_2} \right) X_{n_1n_2}.
\end{equation}
In the special case of Hermitian matrix $X_{n_1n_2}=X_{n_2n_1}^*$, it is shown that $\delta X$ becomes real number by adding the expression (\ref{eq:fluctuation0}) with the subscripts $n_1$ and $n_2$ exchanged.
The order of the fluctuation for a general matrix $X$ is estimated as 
\begin{eqnarray}
\label{eq:fluctuation2}
\left| \delta X \right|^2 &=& 
\frac{1}{K^2} \sum_{k,k'} \sum_{n_1n_2n_3n_4} 
\left( \xi_{n_1}^{*(k)} \xi_{n_2}^{(k)}  -\delta_{n_1n_2} \right) X_{n_1n_2} \nonumber \\
&& \times \left( \xi_{n_3}^{(k')} \xi_{n_4}^{*(k')}  -\delta_{n_3n_4} \right) X_{n_3n_4}^* \\
&=& 
\frac{1}{K^2} \sum_{k,k'} \sum_{n_1n_2n_3n_4} 
\xi_{n_1}^{*(k)} \xi_{n_2}^{(k)} \xi_{n_3}^{(k')} \xi_{n_4}^{*(k')} X_{n_1n_2}X_{n_3n_4}^* \nonumber \\
&+& 
\frac{1}{K} \sum_{k} \sum_{n_1n_2n_3n_4} 
\xi_{n_1}^{*(k)} \xi_{n_2}^{(k)} \left( -\delta_{n_3n_4} \right) X_{n_1n_2}X_{n_3n_4}^* \nonumber \\
&+& 
\frac{1}{K} \sum_{k} \sum_{n_1n_2n_3n_4} 
\left( -\delta_{n_1n_2} \right) \xi_{n_3}^{(k)} \xi_{n_4}^{*(k)}  X_{n_1n_2}X_{n_3n_4}^* \nonumber \\
&+&
\sum_{n_1n_2n_3n_4} \delta_{n_1n_2}\delta_{n_3n_4} X_{n_1n_2}X_{n_3n_4}^* \nonumber
.
\end{eqnarray}
By taking statistical average of (\ref{eq:fluctuation2}) and carefully evaluating $\langle\langle \xi_{n_1}^{*(k)} \xi_{n_2}^{(k)} \xi_{n_3}^{(k')} \xi_{n_4}^{*(k')} \rangle\rangle$ by using (\ref{eq:statistical1})-(\ref{eq:statistical3}), we obtain
\begin{equation}
\label{eq:fluctuation}
\left| \delta X \right|^2 =
\frac{1}{K} \left\{
\left( \langle\langle  |\xi_n|^4 \rangle\rangle -1\right) \sum_n |X_{nn}|^2
+ \sum_{n_1 \ne n_2} |X_{n_1n_2}|^2
\right\}
\end{equation}
where the factor $\left( \langle\langle   |\xi_n|^4 \rangle\rangle -1\right)$ is factored out because  we assumed identical probability distribution for all $\xi_n$.
The factor $1/K$ ensures the behavior $|\delta X| \sim 1/\sqrt{K}$ expected from the central limit theorem.
According to the inequality 
\begin{equation}
\left\langle \left\langle \   |\xi_n|^4 \  \right\rangle \right\rangle
\ge
\left( \left\langle \left\langle \  |\xi_n|^2 \  \right\rangle \right\rangle \right)^2
= 1
,
\end{equation}
the fluctuation becomes the smallest for a given basis set if and only if $ |\xi_n|=1$ for each $n$ and for each sample of random variables. 
One of such random vectors is called {\em random phase vector} \cite{Alben1975,Iitaka1995b,Iitaka1997} and defined by
\begin{equation}
\label{random_eq:randomphasevec}
|\Phi_{random \ phase} \rangle \equiv \sum_{n=1}^N |n \rangle e^{i\theta_n}
\end{equation}
where $\theta_n$ are a set of independent uniform random variables defined in $[-\pi,\pi]$.
Obviously, the random variables $\xi_n=e^{i\theta_n}$ satisfies the statistical relations 
(\ref{eq:statistical1})-(\ref{eq:statistical3}). 
Further, each sample of random phase vectors is automatically normalized without statistical fluctuation,
\begin{eqnarray}
\label{random_eq:normalize_randomphase}
&&\langle \Phi_{random \ phase}| \Phi_{random \ phase} \rangle 
  \nonumber \\
&=& \sum_{n',n} e^{i(\theta_n-\theta_n')} \langle n'| n \rangle 
= \sum_{n=1}^N 1 =N 
.
\end{eqnarray}
This result is consistent with the observation in \cite{Hams2000} that normalized random vectors give less statistical error than unnormalized random vectors.
It is important to notice that the definition of random phase vector, $|\xi_n|=1$, depends on the choice of basis set. For example, let us examine a unitary transformation of a random phase vector $\xi_i=e^{i\theta_i}, \ (i=1,2)$ 
\begin{eqnarray}
\zeta_1 &=& \frac{1}{\sqrt{2}} \xi_1 + \frac{1}{\sqrt{2}} \xi_2 \\
\zeta_2 &=& \frac{1}{\sqrt{2}} \xi_1 - \frac{1}{\sqrt{2}} \xi_2
.
\end{eqnarray}
The transformed random vector $\zeta_n$ of course satisfies the relation (\ref{eq:statistical1})-(\ref{eq:statistical3}), but $|\zeta_n|=1 \pm \cos(\theta_1-\theta_2)\ne 1$. Therefore the random phase vector in the original basis set is not a random phase vector in the new basis set.
For random phase vectors in a given basis set, the fluctuation (\ref{eq:fluctuation}) reduces to a simple form \cite{Iitaka1995b},
\begin{equation}
\label{eq:fluctuation_randomphase}
\left| \delta X \right|^2 =
\frac{1}{K} \sum_{n_1 \ne n_2} |X_{n_1n_2}|^2
\end{equation}
which becomes zero for diagonal matrices as expected.
Note that the fluctuation (\ref{eq:fluctuation_randomphase}) depends on the choice of basis set and that it is very important for reducing fluctuation to choose a basis set that makes off-diagonal matrix elements smallest.
For a Hermitian matrix $X$, in theory, we can choose a basis set that diagonalizes $X$ and remove the fluctuation completely.

As a complex random vector (\ref{random_eq:randomvec}), a real random vector can be defined by using real random variables with identical probability distribution satisfying the statistical relations
\begin{eqnarray}
\label{eq:statistical1real}
\left\langle \left\langle \   \xi_{n} \  \right\rangle \right\rangle &=& 0 \\
\label{eq:statistical2real}
\left\langle \left\langle \  \xi_{n_1}  \xi_{n_2} \  \right\rangle \right\rangle &=& \delta_{n_1n_2}.
\end{eqnarray}
Unlike a complex random vector, a real random vector in one basis set is not necessarily mapped by unitary formation (\ref{random_eq:unitary}) to a real random vector in another basis set. However, if we stick to the original basis set, the equations (\ref{random_eq:randomvector.normal})-(\ref{eq:fluctuation2}) are also valid for real random vectors. Before evaluating $|\delta X|^2$ for real random vectors, let us assume that $X$ is {\em symmetric} matrix for simplicity. This does not limit the generality of our argument because any matrix $X$ can be decomposed into the sum of symmetric and antisymmetric parts,
\begin{equation}
X=\frac{1}{2}(X+X^t) + \frac{1}{2}(X-X^t)=S+A
\end{equation}
where $X^t$ represents the transpose of $X$, 
and ${\rm tr} [X]= {\rm tr} [S]$ since ${\rm tr} [A]=0$. Therefore if $X$ is not symmetric then we may calculate the trace of the symmetric part $S$ in place of $X$.
By taking statistical average of (\ref{eq:fluctuation2}) and carefully evaluating $\langle\langle \xi_{n_1}^{(k)} \xi_{n_2}^{(k)} \xi_{n_3}^{(k')} \xi_{n_4}^{(k')} \rangle\rangle$ by using (\ref{eq:statistical1real})-(\ref{eq:statistical2real}), we obtain
\begin{equation}
\label{eq:fluctuation_real}
\left| \delta X \right|^2 =
\frac{1}{K} \left\{
\left( \langle\langle  |\xi_n|^4 \rangle\rangle -1\right) \sum_n |X_{nn}|^2
+ 2 \sum_{n_1 \ne n_2} |X_{n_1n_2}|^2
\right\}
\end{equation}
As in the case of complex random vectors, the real random vectors with minimum fluctuation are random vectors with $\xi_n=\pm 1$, ({\em random sign vector}).  Random sign vector may be regarded as a random phase vector with binary phase $\theta_n=0, \  \pi$ and satisfies the normalization (\ref{random_eq:normalize_randomphase}) without fluctuation.
For random sign vectors in a given basis set, the fluctuation (\ref{eq:fluctuation_real}) reduces to a simple form 
\begin{equation}
\label{eq:fluctuation_randomsign}
\left| \delta X \right|^2 =
\frac{2}{K} \sum_{n_1 \ne n_2} |X_{n_1n_2}|^2
\end{equation}
which is twice of the fluctuation of random phase vectors (\ref{eq:fluctuation_randomphase}).
Therefore use of random phase vector rather than random sign vector is recommended except in special cases where the evaluation of matrix elements is substantially accelerated by using real numbers instead of complex numbers.

Let us illustrate efficiency of various types of random vectors in the case of the Hamiltonian operator for a particle moving in one-dimensional space under the influence of local potential $V(x)$,
\begin{equation}
H=\frac{p_x^2}{2m}+V(x)
.
\end{equation}
Discretizing the one-dimensional space of size $L=N\Delta x$ into $N$ meshes $x_i= i \Delta x, \ (i=1,\cdots,N)$ by finite difference method gives the matrix of the Hamiltonian, which is tridiagonal,
\begin{equation}
H_{ij}=\frac{\hbar}{2m(\Delta x)^2} \left( \delta_{i,j+1} -2\delta_{i,j}+\delta_{i,j-1}\right)
+\delta_{i,j} V(x_i)
.
\end{equation}
The fluctuation $|\delta H|^2$ for random phase vector, complex random Gaussian vector, random sign vector, and real Gaussian vector is respectively estimated as 2, 6, 4, and 12 in the unit of $\displaystyle \left( \frac{\hbar}{2m(\Delta x)^2} \right)^2$ by using the fact that $\langle\langle  |\xi_n|^4 \rangle\rangle =1, 2, 1, \ {\rm and } \ 3$ respectively and the equations (\ref{eq:fluctuation}) and (\ref{eq:fluctuation_real}).
The fluctuation of random phase vector is 6 times less than that of real Gaussian random vector, which means that 6 times less sampling is necessary for a given accuracy.

In contrast to the random vector dependence of fluctuation, it is rather difficult to discuss the self-averaging effect in a general way because the effect depends on the choice of basis set and also on the physical nature of the matrix $X$ such as whether the quantity is intensive or extensive.  
Therefore let us first examine self-averaging effect of energy density in the above example.
The fluctuation for random phase vector becomes
\begin{equation}
\delta H/L \sim \left( \frac{\hbar}{2m(\Delta x)^3} \right) \frac{\sqrt{2}}{\sqrt{KN}} 
\end{equation}
as $N \rightarrow \infty$, indicating self-averaging behavior $1/\sqrt{N}$.
It is interesting that the local potential $V(x)$ does not contribute to the fluctuation at all in the calculation with the real space basis set.
When the spectral density such as DOS and linear response function, the dimension $N$ of matrix in the above estimation should be replaced by $N_{eff}$, the number of resonances with in the spectral resolution $\eta$, e.g. $N_{eff}=\rho(\omega)\eta$ in case of DOS. Therefore to reach the same accuracy we need more random vectors for higher energy resolution or low temperatures. See Ref.~\cite{Hams2000} for more sophisticated analysis.
To understand general tendency of self-averaging, let us assume typical matrix elements of $X$ have value of O(1). Then the fluctuation $|\delta X|^2$ in (\ref{eq:fluctuation}) and (\ref{eq:fluctuation_real}) becomes $O(N)$ for sparse or banded matrices and $O(N^2)$ for dense matrices. Since the average value, ${\rm tr} [X]$, becomes $O(N)$, the relative fluctuation $\delta X/{\rm tr} [X]$ becomes $O(1/\sqrt{N})$ and $O(1)$, respectively. Therefore self-averaging is effective for sparse or banded matrices but not for dense matrices.

In summary, we have proved that random phase vector is the most efficient choice among random vectors with identical probability distributions satisfying (\ref{eq:statistical1})-(\ref{eq:statistical3}). The fluctuation for random phase vector is expressed as the sum of the square norm of off-diagonal elements. We show also that the {\em smallness} and {\em sparseness} of the off-diagonal elements is important for the self-averaging effect.  Therefore we conclude that it is important for efficient calculation to use random phase vector and to choose a basis set that makes the off-diagonal elements sparse (or banded) and small.

\begin{acknowledgments}
One of the authors (T.I.) thanks S.~Nomura, T.~Tanaka, and M.~Machida for useful discussions.
\end{acknowledgments}

\bibliography{all}

\end{document}